\documentclass[fleqn,usenatbib]{mnras}

\usepackage{newtxtext,newtxmath}

\usepackage[T1]{fontenc}

\DeclareRobustCommand{\VAN}[3]{#2}
\let\VANthebibliography\thebibliography
\def\thebibliography{\DeclareRobustCommand{\VAN}[3]{##3}\VANthebibliography}

\usepackage{graphicx}	
\usepackage{amsmath}	
\usepackage{hyperref}

\newcommand{\mgtwentyfour}{$^{24}\rm{Mg}$}
\newcommand{\mgtwentyfive}{$^{25}\rm{Mg}$}
\newcommand{\mgtwentysix}{$^{26}\rm{Mg}$}
\newcommand{\twentyfivetwentyfour}{$^{25}\rm{Mg}/^{24}\rm{Mg}$}
\newcommand{\twentysixtwentyfour}{$^{26}\rm{Mg}/^{24}\rm{Mg}$}
 
\newcommand{\logg}{$\log g$} 
\newcommand{\kms}{km s$^{-1}$}

\defcitealias{Nissen_Schuster2010}{NS10}
\defcitealias{Kobayashi_Karakas_Lugaro2020}{K20}
\defcitealias{Limongi_Chieffi2018}{LC18}

\title[Mg isotope ratios in an accreted dwarf galaxy]{Magnesium isotope ratios in Milky Way and dwarf galaxy stars}

\author[M. McKenzie et al.]{M. McKenzie,$^{1,2}$\thanks{E-mail: madeleine.mckenzie@anu.edu.au}
S. Monty,$^{3}$
D. Yong,$^{1,2}$
C. Kobayashi,$^{4}$
A. I. Karakas,$^{2,5}$
P. E. Nissen,$^{6}$
J. E. Norris,$^{1}$\newauthor
A. Rains,$^{7}$
A. Mura-Guzmán,$^{1,2}$
E. X. Wang,$^{1,2}$
S. Martell$^{8,2}$\\
$^{1}$Research School of Astronomy \& Astrophysics, Australian National University, Canberra, ACT 2611, Australia, \\
$^{2}$ARC Centre of Excellence for Astrophysics in Three Dimensions (ASTRO-3D), Canberra 2611, Australia,\\
$^{3}$Institute of Astronomy, University of Cambridge, Madingley Road, Cambridge, CB3 0HA, UK, \\
$^{4}$ Centre for Astrophysics Research, Department of Physics, Astronomy and Mathematics, University of Hertfordshire, Hatfield AL10 9AB, UK \\
$^{5}$School of Physics \& Astronomy, Monash University, Clayton VIC 3800, Australia,\\
$^{6}$Department of Physics and Astronomy, Aarhus University, Ny Munkegade 120, DK–8000 Aarhus C, Denmark,\\
$^{7}$Department of Physics and Astronomy, Uppsala University, Box 516,     SE-75120 Uppsala, Sweden,\\
$^{8}$School of Physics, University of New South Wales, Sydney, NSW 2052. Australia\\
}
\date{Accepted 2024 July 25. Received 2024 July 4; in original form 2024 April 14}

\pubyear{2024}

\begin{document}
\label{firstpage}
\pagerange{\pageref{firstpage}--\pageref{lastpage}}
\maketitle

\begin{abstract}

Under the assumption of hierarchical galaxy formation, dwarf galaxies are the closest existing analogues to the high-redshift protogalaxies that merged to form the Milky Way. These low-mass systems serve as unique laboratories for studying nucleosynthetic channels given that the chemical compositions of their stars play a pivotal role in constraining their chemical enrichment history. To date, stellar abundances in dwarf galaxies have focused almost exclusively on elemental abundance ratios. While important, elemental abundances omit critical information about the isotopic composition. Here we compute the Mg isotopic ratios of six accreted dwarf galaxy stars (low-$\alpha$) and seven Milky Way stars (high-$\alpha$) using a set of high-resolution (65,000 < R < 160,000), high-signal-to-noise ratio ($\rm{S/N} > 250$) optical spectra. We show, for the first time, that at a given [Fe/H] stars born in a dwarf galaxy differ in their Mg isotopic ratios from stars born in the Milky Way. However, when comparing isotopic ratios at a given [Mg/H] rather than [Fe/H], a powerful diagnostic emerges that suggests nucleosynthesis processes are consistent across different stellar environments. This universality of Mg isotopic abundances provides additional dimensionality for chemical evolution models and helps to constrain massive star nucleosynthesis across cosmic time.

\end{abstract}




\section{Introduction}

Although small in mass, dwarf galaxies have had an enormous impact on our understanding of the Universe.
As the building blocks of galaxies we see today and closest existing analogues to high redshift proto-galaxies, dwarf galaxies play a critical role in understanding the grand narrative of cosmic history and hierarchical formation \cite[e.g.,][]{Madau+1996, Mateo1998, Moore+1999, McConnachie2012, Bullock_Boylan-Kolchin_2017}.

$\Lambda$CDM cosmology predicts that galaxies grow through a series of merger events, as shown in cosmological simulations \cite[e.g.,][]{Springel+2005, Vogelsberger+2014, Schaye+2015}. Within our own Galaxy, significant progress has been made in characterising the various components of the accreted halo of the Milky Way \cite[MW, e.g.,][]{Belokurov+2018, Helmi+2018, Haywood+2018, Myeong+2019, Matsuno+2019, Monty+2020, Yuan+2020, Naidu+2021, Horta+2021, Feuillet+2021, Horta+2023, Dodd+2023}.
The ongoing disruption of the Sagittarius dwarf spheroidal galaxy \cite[][]{Ibata1994, Davies+2023} and the large number of tidal streams \cite[][]{Helmi+1999, Belokurov+2006, Shipp+2018, Ji+2020} reinforce the hierarchical nature of galaxy formation and the key role dwarf galaxies play in this process. 

Chemical abundance ratios provide a valuable tool for understanding the star formation history (SFH) of a galaxy \citep[][]{Wallerstein_1962, Tinsley_1979, Bensby+2005}. In particular, the mean [$\alpha$/Fe] ratio\footnote{Where $\alpha$ represents elements produced via the $\alpha$ process such as O, Mg and Si.} assesses the contribution from massive stars with short lifetimes which die as core-collapse supernovae (CCSNe), largely producing $\alpha$ elements with modest amounts of Fe, and longer-lived thermonuclear supernovae (SNe Ia) which dominate the production of Fe-peak elements \cite[][]{Matteucci_Greggio1986, Woosley_Weaver1995, Thielemann+1996, Kobayashi+2020, Kobayashi_Karakas_Lugaro2020}.

Additionally, the evolution of [$\alpha$/Fe] with metallicity below the $\alpha$-knee reveals insights into pre-SNe Ia star formation efficiency. In the $\rm{[Fe/H]} >-3$ regime, dwarf galaxies typically show lower $\alpha$ abundances compared to the MW, likely due to differing SFHs \citep[][]{Venn+2004, Tolstoy+2009}.
Galactic chemical evolution models have been used to interpret SFHs in the MW and Local Group dwarf galaxies \citep[e.g.,][]{Kobayashi+2020, Hasselquist+2021}. However, these models are sensitive to predictions of chemical yields from the various nucleosynthetic channels \citep{Cote+2017, Romano+2010}.

\citet[][hereafter, \citetalias{Nissen_Schuster2010}]{Nissen_Schuster2010, Nissen_Schuster2011} identified high and low [$\alpha$/Fe] populations of halo stars in the solar neighbourhood and attributed the appearance of the two sequences to the existence of in-situ and accreted components in the MW halo. Specifically, the high-$\alpha$ population is indicative of efficient star formation in the early MW, while the low-$\alpha$ population was likely accreted from dwarf galaxies that experienced less efficient star formation.
Today, the majority of the low-$\alpha$ component is attributed to the Gaia-Sausage-Enceladus (GSE) accretion event \cite[][]{Helmi+2018,Belokurov+2018, Haywood+2018}, presumed to have occurred $\sim10~\rm{Gyr}$ ago and dominates the population of halo stars below [Fe/H] $<-1.5$. 
The high-$\alpha$ sequence discovered in \citetalias{Nissen_Schuster2010} is likely connected to the population of ``Aurora'' and ``Splash’’ stars discussed in \citet{Belokurov+2020} \citep[see also][]{Bonaca+2017, Gallart+2019}. Aurora is likely an ancient portion of the MW that existed prior to the formation of the disc and has low metallicities and low tangential velocities whereas Splash stars can be found at higher metallicities than both GSE and Aurora but display a range of tangential velocities in addition to a net rotation.

The bright accreted stars from the \citetalias{Nissen_Schuster2010} sample offer an unparalleled opportunity to not only quantify chemical compositions in terms of overall abundances but also understand how the elements are distributed among their constituent isotopes. Although `isotopic analysis is one of the most difficult sub-fields of spectroscopic astronomy' \citep{Shetrone1996}, it offer the most direct insight into nucleosynthesis and chemical enrichment \cite[e.g.,][]{Kobayashi+2011}. Applications of isotopic measurements include CNO isotopic ratios for the interstellar medium \citep[e.g.,][]{romano19}, C isotopic ratios in stars \citep[e.g.,][]{spite06}, and Li isotopic ratios to probe big bang nucleosynthesis \citep[e.g.,][]{Fields_2011, Wang+2022}.

The relative abundances of \mgtwentyfour, \mgtwentyfive\ and \mgtwentysix\ can be quantified via asymmetries in MgH molecular lines at around 5100 \AA{}. The dominant $\alpha$-nucleus $^{24}\rm{Mg}$ is synthesized during C and Ne burning in massive stars. Conversely, the production of the rarer neutron-rich isotopes $^{25}\rm{Mg}$ and $^{26}\rm{Mg}$ occurs during helium burning in massive stars ($\gtrsim 8~\rm{M}_{\odot}$) as well as during helium-shell burning in intermediate mass ($3-8~\rm{M}_{\odot}$) asymptotic giant branch (AGB) stars \cite[][]{Karakas_Lattanzio2003}. The isotopic yields from CCSNe are also metallicity dependent \citep{Kobayashi+2011} and potentially offer a more refined SFH `clock' than elemental abundances alone. Unfortunately, even with efficient modern spectrographs on 8-10 m class telescopes, the exposure times required for Mg isotopic analysis in surviving dwarf galaxy stars are impractical.

Pioneering studies of Mg isotopes iteratively fit synthetic spectra to individual MgH lines by hand \cite[][]{Boesgaard1968, Bell_Branch1970, Tomkin_Lambert1976, Tomkin_Lambert_1980, Barbuy1985, Barbuy1987, Barbuy_Spite_Spite1987, McWilliam_Lambert1988, Gay_Lambert2000}. Subsequent studies improved upon this by applying a grid-based search for the best-fit spectra over a small number of parameters to recover the isotopic ratios \cite[][]{Yong+2003_NGC6752, Yong+2003, Yong+2004, Yong+2006, Melendez_Cohen2007, Yong+2008, Melendez_Cohen2009, DaCosta+2013, Thygesen+2016, Carlos+2018}. Recently, \cite{McKenzie+2023} developed a powerful new tool capable of determining isotopic ratios based on more than three times the number of MgH lines previously used. This tool also takes several additional parameters into account, such as macroturbulent velocity and continuum placement, and assigns realistic errors to the isotopic ratios for the first time.

In this \textit{letter}, we present measurements of Mg isotope ratios in low-$\alpha$ accreted dwarf galaxy stars to reveal details of massive star contributions that cannot be learnt from element abundances alone. In Section \ref{sec:Obs_analysis} we describe our observations and analysis, in Section \ref{sec:results} we discuss our results, and in Section \ref{sec:Dis_con} we present our conclusions.

\section{Observations and Analysis}
\label{sec:Obs_analysis}

\begin{table*}
\centering
    \caption{The stellar parameters and details of our target stars. Stellar parameters and abundances of HD 134439, HD 134440 and HD 103095 come from \protect\cite{Reggiani_Melendez2018} and are shifted on to the abundance scale of \protect\citetalias{Nissen_Schuster2010} for both [Fe/H] and [Mg/Fe] ($\rm{[Fe/H]}_{\rm{NS10}}$ and $\rm{[Mg/Fe]}_{\rm{NS10}}$ respectively). We perform our own differential abundance measurements for GJ1064, LHS3780, PLX5805 and PLX2019 and also shift them to match \protect\citetalias{Nissen_Schuster2010}. For G66-22, we have spectra from both the UVES and MIKE instruments.}
    \label{tab:LP_tab}
\begin{tabular}{lccccccccc} \hline
Name & Gaia DR3 ID & $T_{\rm eff}$& $\log g$ & [Fe/H] & $\xi$ & $\rm{[Fe/H]}_{\rm{NS10}}$ & $\rm{[Mg/Fe]}_{\rm{NS10}}$ & Instrument & Reference \\
 & & (K) & (dex) & & (km s$^{-1}$) &  & &  & \\
\hline
G13-38 & 3700341138433848832 &5263 & 4.54 & -0.88 & 0.9 & -0.88 & 0.35 & UVES & This work \\
G18-28 & 2728314787225296000 & 5372 & 4.41 & -0.83 & 1.0 & -0.83 & 0.36 & MIKE & \cite{Fishlock+2017} \\
G66-22 & 1159108770069883136 & 5236 & 4.41 & -0.86 & 0.9 & -0.86 & 0.08 & UVES, & This work, \\
 &  &  &  &  &  &  &  &  MIKE & \cite{Fishlock+2017} \\
G82-05 & 3202470247468181632& 5277 & 4.45 & -0.75 & 0.9 & -0.75 & 0.06 & UVES & This work  \\
GJ1064 & 231113736385994624 & 5136 & 4.54 & -1.02 & 0.4 & -0.96 & 0.44 & HIRES & \cite{Yong+2003} \\
HD103095 & 4034171629042489088 & 5100 & 4.65 & -1.35 & 0.9 & -1.29 & 0.21 & McDonald & \cite{Gay_Lambert2000} \\
HD134439 & 6307374845312759552 & 5084 & 4.66 & -1.43 & 1.2 & -1.37 & 0.07 & UVES &  This work  \\
HD134440 & 6307365499463905536 & 4946 & 4.68 & -1.39 & 1.2 & -1.33 & 0.03 & McDonald & \cite{Gay_Lambert2000} \\
HD222766 & 2439291667485442688 & 5334 & 4.27 & -0.67 & 0.8 & -0.67 & 0.35 & UVES & This work  \\
HD230409 & 4517777421819724416 & 5318 & 4.54 & -0.85 & 1.1 & -0.85 & 0.30 & MIKE & \cite{Fishlock+2017} \\
LHS3780 & 2615957416265170304 & 4908 & 4.74 & -1.38 & 0.8 & -1.31 & 0.39 & HIRES & \cite{Yong+2003} \\
PLX2019 & 3073097998492775808 & 4786 & 4.66 & -1.30 & 0.6 & -1.24 & 0.23 & HIRES & \cite{Yong+2003} \\
PLX5805 & 2853258035861905664 & 4850 & 4.84 & -1.45 & 1.2 & -1.40 & 0.15 & HIRES & \cite{Yong+2003}   \\ \hline
\end{tabular}
\end{table*}

\subsection{Target selection and observations}
\label{sec:targets} 
We analyse 13 targets in this work. Five targets were observed using the UVES instrument at the VLT \cite[][]{Dekker+2000} with image slicer \#3 and a 0.3" slit which provided a spectral resolution of $\rm{R}=110,000$. Four stars were observed using HIRES \cite[][]{Vogt+1994} at the Keck telescopes. The 0.4" slit was used which provided a spectral resolution of $\rm{R} \simeq 90,000$. Three stars were observed using MIKE \cite[][]{Bernstein+2003} at the Magellan telescope. The 0.35” slit was used providing a spectral resolution of $\rm{R}=65,000$. Two stars were taken with permission from \cite{Gay_Lambert2000}, observed using the 2.7 m Harlan J. Smith reflector and its coudé spectrograph at McDonald Observatory with a spectral resolution of $\rm{R}=160,000$. All observations have S/N ranging from 300 to 400 per pixel near the MgH lines at 5140~\AA.

In Table~\ref{tab:LP_tab} we list which instrument each star was observed with and the original reference to the data. Very few metal-poor stars with measurable MgH lines and high-quality spectra exist in the literature. However, we include as many low-$\alpha$ stars with MgH lines with high-resolution (R > 60, 000), and signal-to-noise ($\rm{S/N} \gtrsim 250$) spectra as are available to us.

\subsection{Stellar parameters and chemical abundances}

When available, stellar parameters and total Mg abundances are from \citetalias{Nissen_Schuster2010}. For the remaining stars, we adopted the following approach. \cite{Reggiani_Melendez2018} presented differential stellar parameters for Gmb1830 (HD103095), HD 134439, HD 134440 and HD 163810. HD 163810 is also included in the \citetalias{Nissen_Schuster2010} sample, and therefore we shift the \cite{Reggiani_Melendez2018} stellar parameters and Mg abundances onto the \citetalias{Nissen_Schuster2010} scale ($\rm{[Fe/H]}_{\rm{NS10}}, \rm{[Mg/Fe]}_{\rm{NS10}}\ $). For the Keck observations, we analysed those stars differentially with respect to HD103095 and thus onto the \citetalias{Nissen_Schuster2010} scale. We provide these stellar parameters and abundances in Table~\ref{tab:LP_tab}.

\subsection{Mg isotopic analysis}

\begin{figure}
    \includegraphics[width=\columnwidth]{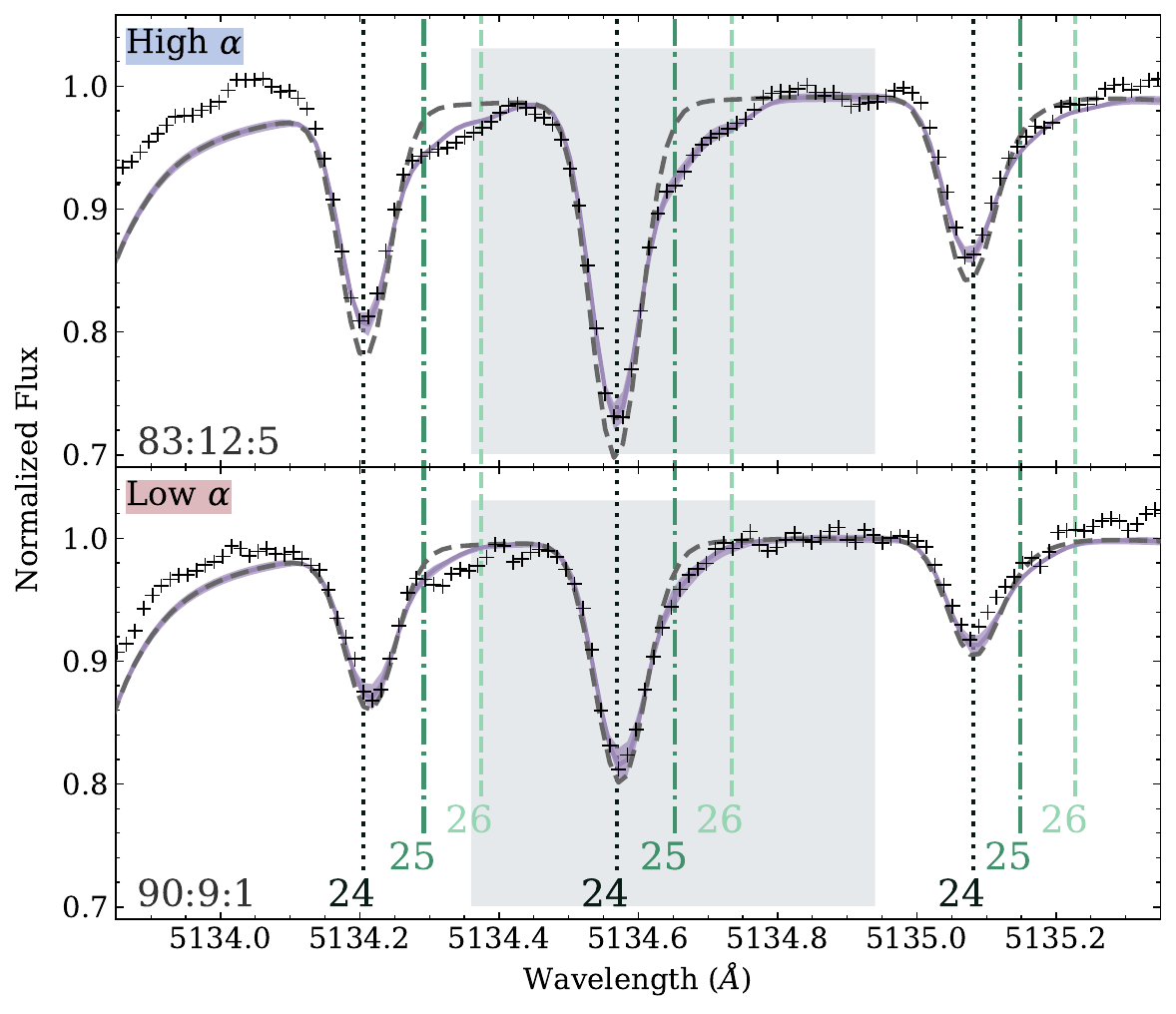}
    \caption{Two stars with similar stellar parameters both observed by the UVES instrument on the VLT; G13-38 (high-$\alpha$) and G66-22 (low-$\alpha$). The black `+' markers represent the normalised spectra, the purple line is the best-fitting model and the grey dashed line is the best-fitting model but with only \mgtwentyfour. The green vertical lines represent the location of the \mgtwentyfour \ (dotted), \mgtwentyfive \ (dash-dotted), and \mgtwentysix \ (dashed) absorption features. The bottom left corner of each panel gives the ratios for \mgtwentyfour:\mgtwentyfive:\mgtwentysix. The high $\alpha$ star has a larger contribution from \mgtwentyfive \ and \mgtwentysix, and has a deeper Mg line.}
    \label{fig:spectra_plot}
\end{figure}

We use the same method outlined in \cite{McKenzie+2023} using our Mg isotopic analysis code \textsc{ratio}\footnote{\url{https://github.com/madeleine-mckenzie/RAtIO}}. The process for determining our isotopic ratios is described in detail in that work. However, we summarise it here for completeness. 

We perform a routine based on Markov chain Monte Carlo optimization of the total Mg abundance, macroturbulent velocity, the ratios of \twentyfivetwentyfour\ and \twentysixtwentyfour, the placement of the continuum and the radial velocity correction. This acts as a wrapper for the code MOOG \cite[][]{Sneden1973, Sobeck+2011} and we use the 2017 version with scattering\footnote{Available from \url{https://github.com/alexji/moog17scat}}.

We give an example of the fits to the 5134.6 \AA{} line (R1 as defined in \citealt{McKenzie+2023}) for two stars observed with UVES; G13-38 (a high-$\alpha$ MW star) and G66-22 (a low-$\alpha$ GSE star) in Fig. \ref{fig:spectra_plot}. The grey rectangle represents the $\chi^2$ fitting region, and the purple line and shaded regions represent the best-fit model and corresponding $16^{\rm{th}}$ and $84^{\rm{th}}$ confidence intervals respectively. The grey dashed line gives the best fitting model if there was only a contribution from \mgtwentyfour\ and no \mgtwentyfive\ or \mgtwentysix. For our high-$\alpha$ MW star, G13-38, there is a marked difference between the \mgtwentyfour\ only model, and our best-fitting model, illustrating the contribution from heavy Mg isotopes. The difference is far less pronounced in the low-$\alpha$ GSE star G66-22.

In the bottom left corner, we write the best-fitting model isotopic ratio. As we take our overall Mg isotopic ratio based on the posterior distributions of our fitting routine, the ratio for the individual line is within the error bounds of the averaged Mg isotope ratio. The neighbouring MgH features 5134.2 \AA{} (R4) and 5135.1 \AA{} (R5) are both well fit by the isotopic ratio found for R1, indicating a good overall fit for each star.

The original application of this code in \cite{McKenzie+2023} analysed metal-poor ([Fe/H] $\approx -1.7$), cool red giant branch stars, rather than the dwarfs with iron abundances between $-1.4 \lesssim \rm{[Fe/H]} \lesssim -0.7$ used in this study. We find that a larger number of MgH lines are well measured for these dwarfs compared to giants (e.g., see Table \ref{tab:isotopes}), leading to more robust estimates of the isotopic ratio.

\subsection{Membership classification}
\label{sec:kinematics}

We interpret our results in the context of the \citet{Belokurov+2020} Aurora-GSE-Splash scenario for the creation of the MW halo.
Membership for our stars is assigned using their integrals of motion and orbital characteristics. We integrate orbits for our sample using the orbital dynamics package \texttt{galpy} \citep{Bovy2015} assuming the \texttt{MWPotential2014} representation of the MW potential \citep[details of the potential components are given in table~1 of][]{Bovy2015}. To orient ourselves, we adopt a circular velocity of 229~\kms\ at the Solar circle \citep{Eilers2019}, assume  $(U,V,W)=[11.1, 12.24, 7.25]$~\kms\ \citep{Schonrich+2010} for the Solar peculiar velocity, 8.121~kpc for the distance to the Galactic centre \citep{Gravity+2018} and assume that the Sun is 20.8~pc above the Galactic plane in the z-direction \citep{Bennett2019+}.

For every star in our sample, we adopt the Gaia DR3 proper motions and radial velocities  \citep[][]{Gaia_DR3} and geometric distances from the \citet{Bailer-Jones+2021} catalogue to build the initial 6D phase-space coordinates for the stars. The orbits are integrated forward for 2~Gyr to determine the maximum height from the disc (z$_{\mathrm{max}}$), pericentric (R$_{\mathrm{peri}}$) and apocentric radii (R$_{\mathrm{apo}}$), eccentricity ($e$), energy and z-component of the angular momentum (L$_{\mathrm{z}}$) directly. The actions are computed using the St$\ddot{\text{a}}$ckel ``Fudge'' method implemented in \texttt{galpy} \citep{Binney2012, Mackereth+2018}. Based on our kinematic measurements, we determine that our sample contains two stars from Aurora, five from Splash, and six from GSE (see Table~\ref{tab:isotopes}).

\section{Results and Discussion}
\label{sec:results}
\subsection{Mg isotopic ratios and kinematic associations}
\begin{table*}
\caption{Isotopic ratios and kinematic associations for our target stars. \# lines refers to the number of lines that were used to measure the isotopic ratios \protect\citep[see][]{McKenzie+2023}. For G66-22 we report isotopic abundances from our UVES spectra. However, these results are consistent with our MIKE spectra within our error bounds.}
\label{tab:isotopes}
\begin{tabular}{lccccccccc}
\hline
Name & \twentyfivetwentyfour & \twentysixtwentyfour & \mgtwentyfour & \mgtwentyfive & \mgtwentysix & \# lines & Component & Energy & L$_\mathrm{z}$ \\
& & & (\%) & (\%) & (\%) & & & ($\mathrm{km^2\,s^{-2}}$) & ($\mathrm{kpc\,km\,s^{-1}}$) \\
\hline
G13-38 & 0.128 ~ ($\pm$0.021) & 0.061 ~ ($\pm$0.016) & 84 ~ ($\pm$2.7) & 11 ~ ($\pm$1.5) & 5 ~ ($\pm$1.2) & 6 & Splash & -57834 & 736\\
G18-28 & 0.158 ~ ($\pm$0.080) & 0.075 ~ ($\pm$0.039) & 81 ~ ($\pm$7.4) & 13 ~ ($\pm$4.8) & 6 ~ ($\pm$2.5) & 7 & Splash & -71841 & -41 \\
G66-22 & 0.063 ~ ($\pm$0.037) & 0.015 ~ ($\pm$0.020) & 93 ~ ($\pm$4.5) & 6 ~ ($\pm$2.9) & 1 ~ ($\pm$1.6) & 5 &  GSE & -30935 & 124\\
G82-05 & 0.062 ~ ($\pm$0.036) & 0.017 ~ ($\pm$0.019) & 93 ~ ($\pm$4.3) & 6 ~ ($\pm$2.8) & 1 ~ ($\pm$1.5) & 6 &  GSE & -10320 & 465\\
GJ1064 & 0.148 ~ ($\pm$0.025) & 0.068 ~ ($\pm$0.010) & 82 ~ ($\pm$2.5) & 12 ~ ($\pm$1.8) & 6 ~ ($\pm$0.7) & 7 &  Splash & -57683 & 1057\\
HD103095 & 0.077 ~ ($\pm$0.013) & 0.008 ~ ($\pm$0.007) & 92 ~ ($\pm$1.7) & 7 ~ ($\pm$1.1) & 1 ~ ($\pm$0.6) & 6 & GSE & -26158 & 672 \\
HD134439 & 0.043 ~ ($\pm$0.013) & 0.008 ~ ($\pm$0.008) & 95 ~ ($\pm$1.9) & 4 ~ ($\pm$1.2) & 1 ~ ($\pm$0.7) & 7 &  GSE & 16051 & -2200\\
HD134440 & 0.064 ~ ($\pm$0.007) & 0.004 ~ ($\pm$0.003) & 94 ~ ($\pm$0.9) & 6 ~ ($\pm$0.6) & 0 ~ ($\pm$0.3) & 7 &  GSE & 15981 & -2202\\
HD222766 & 0.125 ~ ($\pm$0.050) & 0.075 ~ ($\pm$0.030) & 83 ~ ($\pm$5.9) & 11 ~ ($\pm$3.7) & 6 ~ ($\pm$2.3) & 5 &  Splash & -54867 & 403\\
HD230409 & 0.081 ~ ($\pm$0.055) & 0.051 ~ ($\pm$0.027) & 88 ~ ($\pm$5.8) & 7 ~ ($\pm$3.9) & 4 ~ ($\pm$1.9) & 6 &  Splash & -51788 & 945\\
LHS3780 & 0.070 ~ ($\pm$0.010) & 0.016 ~ ($\pm$0.003) & 92 ~ ($\pm$1.1) & 6 ~ ($\pm$0.8) & 2 ~ ($\pm$0.3) & 5 &  Aurora & -62286 & 334\\
PLX2019 & 0.056 ~ ($\pm$0.008) & 0.008 ~ ($\pm$0.004) & 94 ~ ($\pm$0.9) & 5 ~ ($\pm$0.7) & 1 ~ ($\pm$0.2) & 5 &  GSE & 	-42061 & -597\\
PLX5805 & 0.057 ~ ($\pm$0.007) & 0.007 ~ ($\pm$0.003) & 94 ~ ($\pm$0.9) & 5 ~ ($\pm$0.6) & 1 ~ ($\pm$0.2) & 6 & Aurora & -46537 & 1079\\
\hline
\end{tabular}
\end{table*}

\begin{figure*}
    \includegraphics[width=\textwidth]{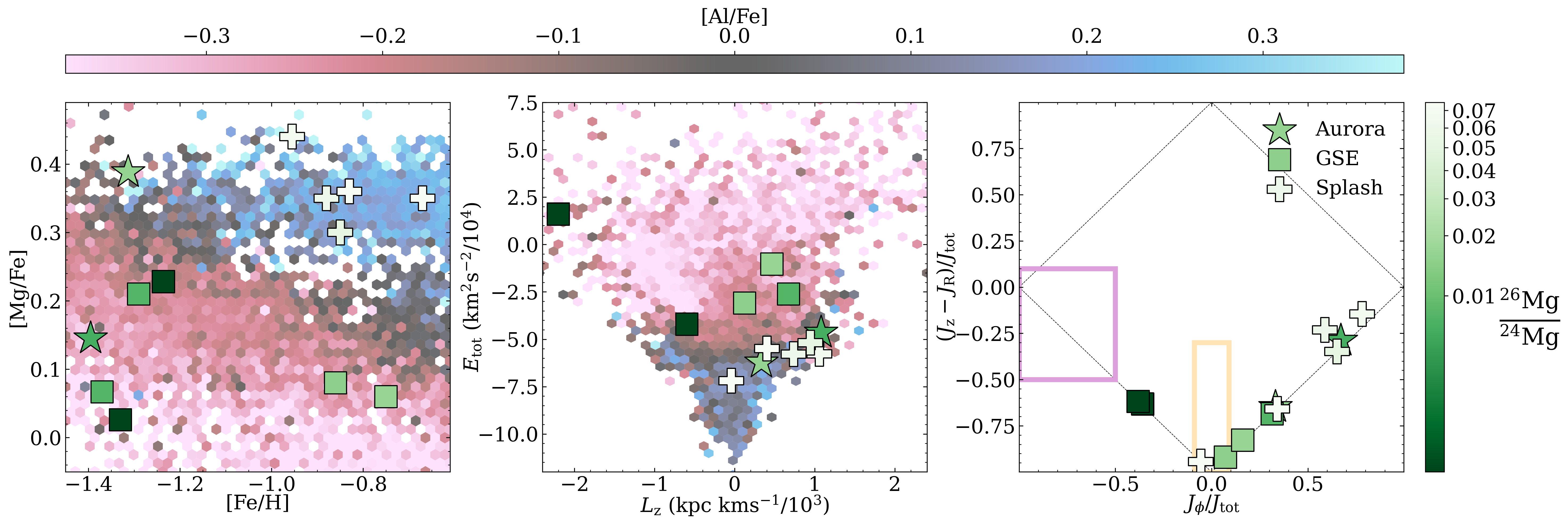}
    \caption{Membership classifications coupled with the Mg isotope ratio \twentysixtwentyfour \ in terms of [Mg/Fe] against [Fe/H] (left), energy and angular momentum (centre) and action space (right). Targets belonging to the Aurora component are given as a star marker, Splash as a plus marker, and GSE as a square. The leftmost and central panels use APOGEE data binned using the \texttt{matplotlib.hexbin} function to contextualise our membership selection. In the rightmost panel, the orange rectangle at the bottom, and purple rectangle to the left represent the approximate locations of the GSE and Gaia-Sequoia accretion events as given by \protect\cite{Myeong+2019} and as shown in \protect\cite{Monty+2020}. We use a logarithmic colour bar for our Mg isotope ratio to better distinguish between stars with a small contribution from \mgtwentysix. HD134440 and HD134439 have very similar orbital parameters and sit on top of each other in the energy and angular momentum, and action space plots.}
    \label{fig:kinematics_plot}
\end{figure*}

\begin{figure*}
    \includegraphics[width=\textwidth]{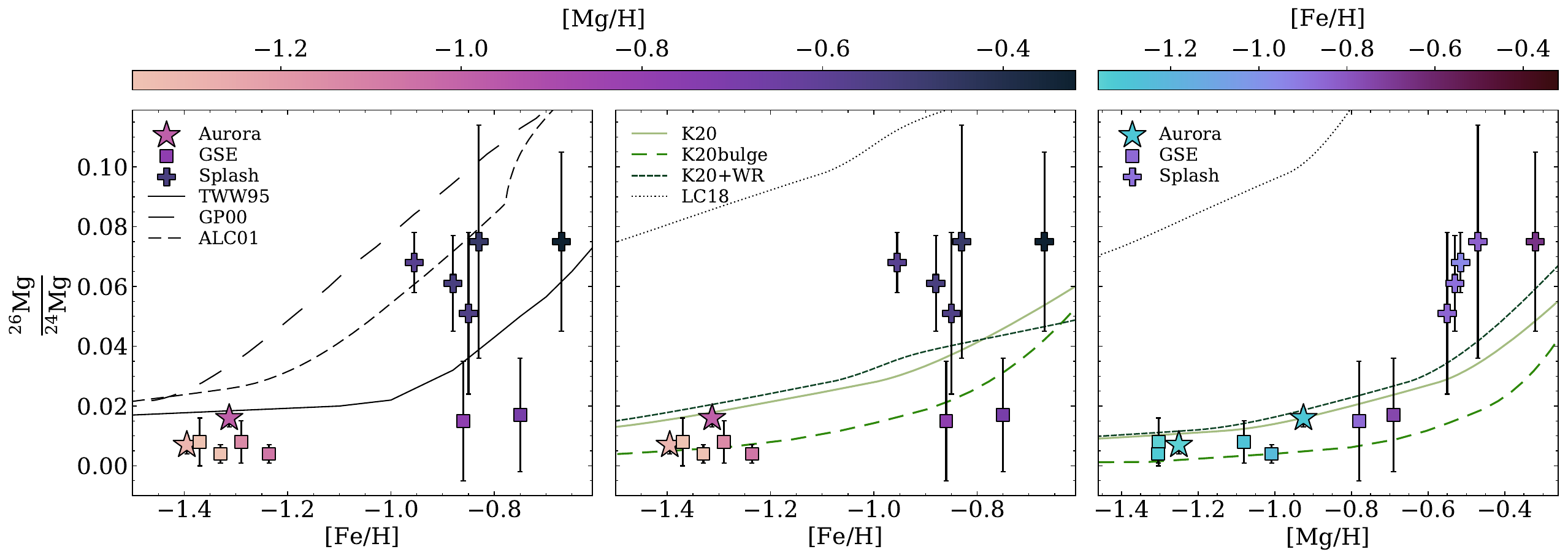}
    \caption{The correlation between \twentysixtwentyfour \  and [Fe/H] (left, and centre), and [Mg/H] (right). We use the same markers in Fig. \ref{fig:kinematics_plot} to represent the different MW components. In the leftmost panel, we plot the chemical evolution models from \protect\citet[][TWW95, solid line]{Timmes_Woosley_Weaver1995}, \protect\citet[][GP00, long-dashed line]{Goswami_Prantzos2000} and \protect\citet[][ALC01, short-dashed line]{Alibe2001} as given in \protect\cite{Yong+2003}. In the central and rightmost panels, we plot models from \protect\citet[][]{Kobayashi+2000, Kobayashi_Karakas_Lugaro2020} for the \citetalias{Kobayashi_Karakas_Lugaro2020} (solar neighbourhood, solid light green line), the  \citetalias{Kobayashi_Karakas_Lugaro2020}bulge (bulge, long-dashed medium green line), the \citetalias{Kobayashi_Karakas_Lugaro2020}+WR (solar neighbourhood with Wolf--Rayet stars, dark green short-dashed line) and the rotating massive stars from \protect\citet[][\citetalias{Limongi_Chieffi2018}, black dotted line]{Limongi_Chieffi2018}. When plotting [Mg/H], the stars seem to follow a universal chemical evolution track, and is best reproduced by both the \citetalias{Kobayashi_Karakas_Lugaro2020} and \citetalias{Kobayashi_Karakas_Lugaro2020}+WR.}
    \label{fig:isotopes}
\end{figure*}

To present our isotopic ratios and contextualize our chemodynamical associations, Fig. \ref{fig:kinematics_plot} features plots of [Mg/Fe] vs. [Fe/H] (left), energy (E) vs. angular momentum (middle), and action space (right). Hereafter, we use star markers to represent members of the Aurora population, plus markers for the Splash population, and squares for GSE. In the leftmost and centre panels, we show a selection of halo giants (\logg < 2.5) from the Apache Point Observatory Galactic Evolution Experiment (APOGEE) DR17 \texttt{allstar-lite} \citep{apogeedr17} catalogue across different metallicities and [Al/Fe]. We choose [Al/Fe] as the third dimension in Fig. \ref{fig:kinematics_plot} as it is known to be underabundant in the accreted halo \citep{Hawkins+2015}, likely a reflection of inefficient star formation in the GSE dwarf.

The \texttt{allstar-lite} catalogue was cleaned to remove all flagged measurements following the methodology of \citet{Belokurov+2022}. To select the halo sample, we keep only stars with absolute $z$-component of the angular momentum $|\mathrm{L}_\mathrm{z}|<=3000$ \kms kpc, z$_\mathrm{max}>3$~kpc and an azimuthal velocity v$_{\phi}=80$~\kms\ using values from the \texttt{astroNN} APOGEE DR17 Value Added Catalogue \citep[][]{Leung2019a+, Leung2019b+}\footnote{The apogee\_astroNN-DR17.fits table is available from \url{https://www.sdss.org/dr18/data_access/value-added-catalogs/?vac_id=85}}. Following this initial cut, we re-determine the values of E and $\mathrm{L}_\mathrm{z}$ for the APOGEE halo sample assuming the same frame of reference as our program stars (described in Sec.~\ref{sec:kinematics}).

We colour our target stars by their isotopic ratio for \twentysixtwentyfour\ using a green colour bar, with a lighter colour representing a larger contribution from the heavier isotope \mgtwentysix. We only show \mgtwentysix\ as it is less sensitive to 3D effects than \mgtwentyfive\ based on 3D hydrodynamical modelling \citep[][]{Thygesen+2016, Thygesen+2017} and because the errors are smaller than for \mgtwentyfive. However, a similar pattern emerges using \mgtwentyfive. A significant result of this \textit{letter} is that at metallicities greater than $\rm{[Fe/H]} = -1$, stars formed within the GSE component have a smaller contribution from \mgtwentysix \ compared to their MW counterparts. 
One explanation is that the MW has experienced more CCSNe than GSE, thus enriching the Galaxy with more heavier Mg isotopes.

\subsection{Mg isotopic ratios and chemical evolution}
Fig. \ref{fig:isotopes} presents \twentysixtwentyfour \ as a function of [Fe/H] and [Mg/H]. In the left panel, we include model predictions from \cite{Timmes_Woosley_Weaver1995}, \cite{Goswami_Prantzos2000} and \cite{Alibe2001}, as in \citet{Yong+2003}. For a description and discussion of these models, we refer the reader to \citet{Yong+2003}. These models overpredict the production of  \twentysixtwentyfour\ at metallicities lower than $\rm{[Fe/H]} = -1.2$, and diverge for metallicities above this. 
In the middle panel, we plot four different models: 1) the solar-neighbourhood and 2) bulge models from \citet[][referred to as \citetalias{Kobayashi_Karakas_Lugaro2020} and \citetalias{Kobayashi_Karakas_Lugaro2020}bulge, respectively]{Kobayashi_Karakas_Lugaro2020}  to demonstrate the dependence of the SFH; 3) the updated solar-neighbourhood models with Wolf--Rayet winds (\citetalias{Kobayashi_Karakas_Lugaro2020}+WR); and 4) models that include rotating massive star yields from \citet[][\citetalias{Limongi_Chieffi2018}]{Limongi_Chieffi2018} assuming the rotational velocity distribution described in \citet{prantzos2018}.
We note that the \citetalias{Kobayashi_Karakas_Lugaro2020}+WR model uses the same yields as in \citet{kobayashi2024} for a high-redshift galaxy, while the \citetalias{Limongi_Chieffi2018} model overproduces [(Ba,Sr,Y,Zr)/Fe] ratios (see fig. 8 of \citealt{kobayashi2022} for `\citetalias{Kobayashi_Karakas_Lugaro2020}sr+rot1' and `\citetalias{Kobayashi_Karakas_Lugaro2020}sr+rot2', respectively).

When comparing to [Fe/H], the bulge model with a rapid star formation from \citetalias{Kobayashi_Karakas_Lugaro2020} does an excellent job in reproducing the low-metallicity end of our measurements, with the solar neighbourhood model and solar-neighborhood models with Wolf--Rayet winds only slightly overestimating \twentysixtwentyfour. This is consistent with the high star formation efficiency found in GSE by \citet{Hasselquist2021+}. The \citetalias{Limongi_Chieffi2018} model overproduces the amount of \mgtwentysix. This is similar to results from \cite{Vangioni_Olive2019} which also see an overestimation of \mgtwentysix \ when adopting \citetalias{Limongi_Chieffi2018} models.

[Mg/H] is used in place of [Fe/H] in some galactic chemical evolution studies \citep[e.g.,][]{Cayrel+2004, Griffith+2019, Griffith+2022, Weinberg+2022, Nissen+2024} and therefore we use this as the x-axis in our rightmost panel, including the same \cite{Kobayashi_Karakas_Lugaro2020} models from the middle panel. Our stars follow a remarkably consistent increase in \twentysixtwentyfour, which is reproduced by all models except for the \citetalias{Limongi_Chieffi2018} model. The Wolf--Rayet model can best reproduce the upturn in \twentysixtwentyfour\ at around $\rm{[Mg/H]}\approx -0.7$, but still underestimates the enhancement in \mgtwentysix. This may support a slight contribution from rotating massive stars at high metallicities.

When considering the right-hand panel of Fig. \ref{fig:isotopes}, it appears as though the \twentysixtwentyfour\ ratio follows a single chemical enrichment track. As described in \cite{Gay_Lambert2000}, the abundance of \mgtwentyfive\ and \mgtwentysix\ increases with metallicity as they are predominantly secondary species with their production relying on the abundance of the seed nucleus $^{22}\rm{Ne}$. 
However, \mgtwentyfive\ and \mgtwentysix\ can occur as a {\it primary} process in AGB stars and rotating massive stars. In massive stars, these heavier Mg isotopes can also be produced from $^{24}\rm{Mg}$ \citep{Kobayashi+2011}. Therefore the minor primary production of \mgtwentyfive\ and \mgtwentysix\ is reflected by our low, but non-zero \twentysixtwentyfour\ isotopic ratios at low [Mg/H]. As [Mg/H] increases, the \twentysixtwentyfour\ ratios rise in accordance with predictions from \cite{Kobayashi_Karakas_Lugaro2020}.

The accreted and in-situ populations appear to follow a similar chemical track suggesting that the evolution of isotopic ratios of Mg is independent of the host galaxy, and is controlled primarily by nucleosynthesis, namely, CCSNe. This hints at a universality of the nucleosynthesis of Mg isotopes in the local Universe across cosmic time. Compared to the isotopic ratios from \cite{Gay_Lambert2000} and \cite{Yong+2003}, our improved measurements have lowered the amount of \mgtwentysix\ at the lower metallicity end of our isotopic ratios, making them more in line with theoretical models. \cite{Fenner+2003} invoked a contribution from AGB stars to explain the scatter and elevated isotopic ratios in these previous measurements. However, the relationship between Mg isotopes and [Mg/H] found by this study indicates that the contribution of AGB stars to the Galactic inventory is small when compared to CCSNe. This is in contrast to work from \cite{Melendez_Cohen2007} who found that the AGB contribution in the Galactic halo increased after $\rm{[Fe/H]} \sim -1.5$. Our results agree with more recent chemical evolution models (see fig.\,18 of \citealt{Kobayashi+2011}) and we conclude that the variation in Mg isotopes against [Fe/H] in the left and middle panels is likely due to the contribution of Fe from SNe Ia.

This interpretation is based on our small sample size, thus illustrating the need for a larger sample of high-resolution ($\rm{R} > 60,000$), high signal-to-noise ($\rm{S/N} \gtrsim 250$) spectra to build a more comprehensive picture of Mg isotope production. Such a sample would help constrain the primary synthesis of heavy Mg isotopes, and define the metallicity at which secondary heavy Mg isotope production becomes an important factor in galactic chemical evolution.

\section{Conclusions}
\label{sec:Dis_con}
For the first time we are able to probe stellar nucleosynthesis at the isotopic level in stars accreted from the GSE dwarf galaxy. We measure Mg isotopic ratios from high resolution and high signal-to-noise spectra for 13 stars using our Mg isotope code \textsc{ratio}, and have assigned the kinematic membership to Aurora, Splash and GSE components. Our results show that MW stars have a larger fraction of \mgtwentyfive \ and \mgtwentysix \ at a given [Fe/H], compared to accreted dwarf galaxy stars. However, a coherent picture of the evolution of the Mg isotopes emerges when we consider the contributions from CCSNe for which the \twentysixtwentyfour\ ratio appears to trace a universal chemical enrichment track. This study therefore demonstrates that while there are differences in [Mg/Fe] at a given [Fe/H] in dwarf galaxies relative to the MW, the isotopic Mg abundances suggest the universality of nucleosynthetic yields across the different star formation environments.

This \textit{letter} encourages 1) further observational campaigns to comprehensively map out the isotope distribution in many more components of the MW, and 2) galactic chemical evolution codes to report not only their total Mg abundances, but the relative abundance of \mgtwentyfour, \mgtwentyfive\ and \mgtwentysix. Our research contributes to a deeper understanding of galactic evolution, emphasizing that key isotopic ratios like \twentysixtwentyfour\ maintain a uniform pattern of enrichment predominantly driven by CCSNe, irrespective of the galactic setting. These insights not only refine our understanding of chemical evolution in different galactic environments but also pave the way for more precise observations and models of galaxy formation and evolution.

\section*{Acknowledgements}
The authors thank the anonymous referee for their helpful comments which improved the clarity of this work. This research was based on observations collected at the European Southern Observatory under ESO programme 0100.D-0072(A). The authors made use of the Keck Observatory Archive (KOA), which is operated by the W. M. Keck Observatory and the NASA Exoplanet Science Institute (NExScI), under contract with the National Aeronautics and Space Administration. The authors thank Prof. Vasily Belokurov for constructive conversations on chemodynamic space and Dr Pamela L. Gay for sharing spectra with us.
This work was supported by the Australian Research Council Centre of Excellence for All Sky Astrophysics in 3 Dimensions (ASTRO 3D), through project number CE170100013.
CK acknowledges funding from the UK Science and Technology Facilities Council through grants ST/R000905/1, ST/V000632/1, ST/Y001443/1, and also the Stromlo Distinguished Visitorship at the ANU. MM acknowledges the traditional custodians of the land on which the Australian National University is based, the Ngunnawal and Ngambri peoples, and pays their respects to elders past and present.
This work makes use of Jupyter notebooks \citep{Kluyver2016jupyter} as well as the Python packages \texttt{NumPy} \citep{harris2020array}, \texttt{Matplotlib} \citep{Hunter2007}, \texttt{Pandas} \citep{reback2020pandas}, and \texttt{emcee} \citep{Foreman-Mackey+2013}.

\section*{Data Availability}
UVES spectra are available from the ESO archive under the program reference 0100.D-0072(A). All other data, including the individual MgH line fits, are available upon reasonable request.



\bibliographystyle{mnras}
\bibliography{main} 




\appendix




\bsp	
\label{lastpage}
\end{document}